# Future Communication Model for High-speed Railway Based on Unmanned Aerial Vehicles

Yuzhe Zhou
School of Electronic & Information Engineering
Beijing Jiaotong University
No.3 Shang Yuan Cun, Haidian District, Beijing, 100044, China
E-mail: 08211052@bjtu.edu.cn

**Abstract**

High-speed railway is playing an important role in mass transportation, due to its lower energy consumption, less environmental pollution, larger capacity and higher safety features. The development of high-speed railway makes people's life more and more convenient. Meanwhile, providing high quality of service broadband communications for fast-moving users still remains unsolved, despite the fact that new solutions of incremental improvements are keeping up with this unprecedented communication requirement growth. This article proposes a communication system infrastructure based on airborne relay for high-speed trains in the further Cyber-Physical Systems. Comparisons and feasibility analysis are provided as well as discussions of key wireless technologies and obstacles in this system.

## 1    Introduction

As the next generation of the complex intelligent system, Cyber-Physical Systems (CPS) [1] is based on the technologies of networks and embedded systems. The CPS stresses the integration of computing, communication, controlling and physical worlds. It has features such as self-adaptability to the environment uncertain change, dynamic reconfiguration and network-based integration control of large-scale systems.

With the advances in communications, embedded systems, sensor and energy storage as well as material technologies enveloped by the CPS concepts, Unmanned Aerial Vehicles (UAVs) became a feasible approach to the CPS. The benefits of UAVs are being realized by potential UAV users outside of military domains and into new areas, such as entertainment, media, surveying, weather forecast and observation, communication relaying, traffic monitoring and management, and even engineering commanding and controlling. Micro- or small- unmanned aircraft systems (UASs) [2] have the potential to create new applications and markets in civil domains. In near future, a large number of UAVs deployed as an efficient medium will be major approaches in realization of the CPS concepts. Aerial ad-hoc networking systems---located just above our heads---should be considered into the future networked communication architectures.

The development of high-speed railways makes people's life more and more convenient. Meanwhile, it puts forward much higher requirements on high-speed railway communication services. As one of the CPS subsystems, Transportation Control System, especially Train Control System, is great protection of the national economy and public safety. Intelligent Transportation System shall meet the CPS characteristics including heterogeneous, autonomous, real-time performance, lower energy consumption, large capacity, robustness, reliability, and safety. For example, China Train Control System Level 3 (CTCS-3) is based on the Global System for Mobile Communication for Railways (GSM-R) which acts as the radio interface between the trains and the control center in order to exchange safety messages. Since broadband and high-speed mobility communications is an unavoidable trend according to the future users' requirements, simply using GSM-R in CTCS-3 does not fulfill these urgent requirements. So, novel technical solutions are desired to keep up with these requests (Fig. 1). Current researches focus on providing fast wireless network access services for high-speed mobile terminals. However, some technical bottlenecks are still standing in the way, and this access issue remains unsettled.

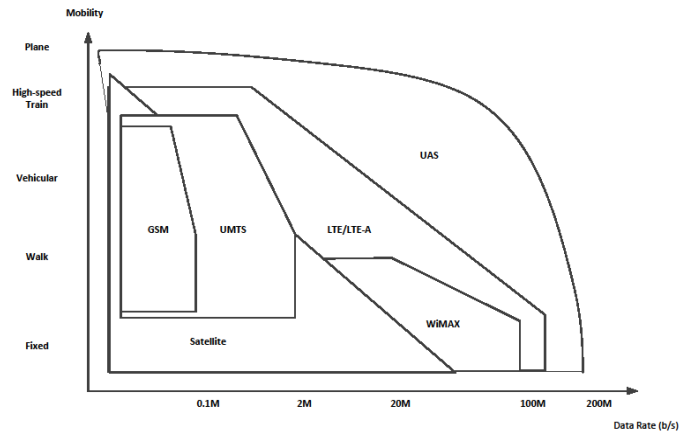

**Fig. 1**  Comparison of current technologies

In this paper, a system that realizes real-time and broadband communications for high-speed train (HST) users is proposed. This system is based on heterogeneous UASs, where UAVs work as communication hot spots or relays along with the moving HST. In this case, a kind of moving cell called Aerial Moving Cells (AMC) is formed, and this system is denoted as UAS-R (Unmanned Aircraft System for Railways). In UAS-R, the ground control station (GCS) for UAVs is replaced by the ground station (GS) which distinguishes from the base station (BS) in the communication terminology. The GS has the control over related UAVs and the capability to forward or receive large data traffic.

This paper is structured as follows. Section 2 illustrates the research motivation of the UAS-R system. The communication architecture of UAS-R is introduced in detail in Section 3. Section 4 discusses the feasibility of the proposed system. Finally, Section 5 summarizes this paper.

## 2  Motivation

The innovation is mainly inspired by the moving base stations concept [3], the moving cells concept [4], and the civil drone amateurs' interests. This paper shifts our attention from the satellite or ground cellular system to the air relay system which is more flexible and efficient.

### 2.1  Satellite communications

Satellite provides sufficient coverage to a large train operation area. Thus, there is no need to install a lot of ground communication equipment next to the tracks. The Doppler Effect almost has no impact on satellite communications. Less handover events occur due to the height of the satellite. However, blind spots exist when the satellite views are obstructed because of terrain or clutter. In addition, satellite communication links are bandwidth limited and have high costs. Therefore, satellite communications cannot provide massive high quality of service (QoS) communications. Besides, satellite communications have considerably connection delay which makes them not suitable for real-time applications.

Compared with the satellite communication platform, UAS-R has a simple structure, high data rate, and low communication hysteresis. Additionally, satellite communications can combine with UAV communications in heterogeneous UASs.

**2.2 Cellular communications**

Cellular infrastructure enables ubiquitous mobile devices to communicate with each other via a lot of fixed BSs. The cellular architecture offers several advantages: 1) Its coverage area is large and can be extended via BSs. 2) Its bandwidth can be reused over different regions which will increase system capacity. 3) The BSs provide redundancy in case that one connection is poor, another connection may be better. However, considering the applications in high-speed railway communications, the cellular architecture encounters some severe problems. For example, frequent handover due to the high-speed moving among cells, handover failure due to Doppler frequency shift, and low data rate caused by fast fading and multipath loss.

Compared with the cellular, a single UAV can cover a larger area and the data rate can achieve a higher level for a specific user. Since the HST running space is very large, a lot of BSs should be deployed to construct a well performed network. However, the BSs are not working all the time, because the HST is only moving within one or two BSs. Therefore, a great initial investment and maintenance costs are needed, and large resources are wasted. In the proposed system, only one UAV is related to one HST for a certain period time or distance. Besides, only a few GSs are necessary in order to meet the communication requirements. It is more efficient and economical compared with cellular networks.

## 2.3 P2P relaying

Mobile peer-to-peer (P2P) network application is limited by the inferior channel condition constraints, which results in reducing communication range and coverage, and high power consumption. Multi-hop P2P network is implemented to overcome these limitations. However, a lot of relay nodes are involved in, and they suffer from long delays and unreliable high-density physical connections.

Compared with omnidirectional antennas in the P2P relay, airborne relay can benefit from the higher gain of a narrow-beam antenna. Because long-hop or short multi-hop P2P relays suffer from high path loss and considerable delay or jitter, the airborne nodes are better for long-range relays. Therefore, the airborne relays are appropriate to reduce coverage loopholes, particularly in mobile dynamic networks.

## 3  UAS-R communication architecture

### 3.1  System model

There are four basic communication architectures that can be used for UAS applications: direct link, satellite, single cellular, and peer-to-peer networking. Each of these architectures has advantages and disadvantages. A direct link that connects the GS and the UAV directly forms the simplest architecture. However, obstructions can block the signal, and the UAV requires a high-power transmitter, a steerable antenna, or significant bandwidth in order to support high-data-rate links for long-range communications. However, the UAV will suffer high delays if their communications are mediated by satellites. So, optional satellite-to-UAV link suits for non-real-time data transmission. Supposing that the coverage area of a UAV is a kind of moving cell, this moving cell provides communications among lots of stationary and mobile user terminals. In the airborne peer-to-peer networking architecture, each UAV acts as a relay node to receive and forward data. Communications between a UAV and a GS or a terminal can take place over several hops through intermediate UAVs. This architecture is similar to a high-altitude cellular architecture [5]. The open space simplifies the connection requirements, and the bandwidth can be reused more frequently and efficiently. Besides, the nodes are mobile in order to support potential communications.

Note that each of these four architectures is not standing alone. All of the involved elements can communicate with each other separately, cooperatively, and simultaneously through unicast, multicast, and broadcast technologies. For the high-speed railway communications, the HSTs are regarded as equivalent mobile terminals. In an AMC, only one HST is associated with this UAV. The HST transmits data to the UAV via a particular designed antenna erected on the roof top of the train body. Wireless signals could feed into the compartment through this antenna, and then reaches the user terminals via distributed antenna systems (DAS) located in the train compartment. The DAS could integrate data from various standards into one traffic stream and differentiate this traffic stream into data of different standards. The GSs are connected to train control center and the Internet providers directly by optical fibers. Therefore, the control data and measurement data can be easily transported to the train control center for operation decisions. Meanwhile, the user data can be directly transported over IP-based networks rather than relayed to other BSs or Mobile Switching Center (MSC) which is the case in current wireless mobile systems. Thus the proposed UAS-R system in Fig. 2 includes HSTs, UAVs, GSs, and other communication facilities involved in. It has the flexibility, robustness, and range-extension features, etc.

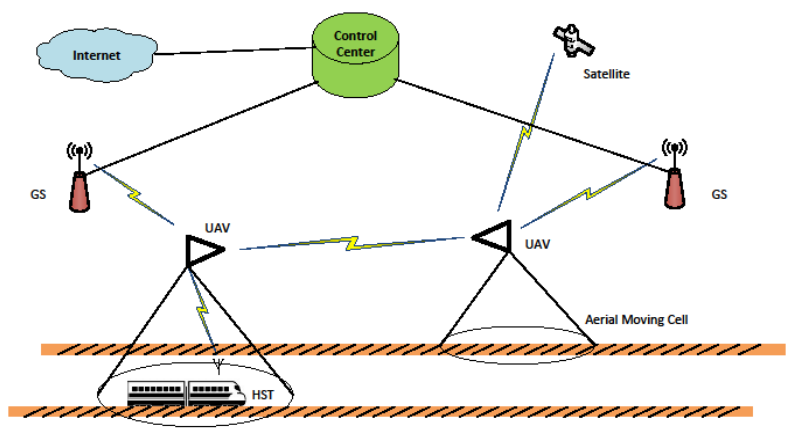

**Fig. 2** The proposed UAS-R system model

## 3.2 Three-hop layer model

The two-hop model [6] introduces a dedicated multi-band and multi-mode base station inside the HST. The first hop is from the BS to the HST, and the second hop is from the HST to the user terminals. This model is modified to a three-hop model in this paper, and the proposed layer model is shown in Fig. 3. The first hop is from the GS to the UAV, the second hop is from the UAV to the HST, and the third hop is from the HST to the user terminals. In the proposed UAS-R system, there is no need to equip a large number of BSs near the trackside. Instead, only a few GSs far away from the tracks are required. The proposed model can significantly mitigate the Doppler Effect, the multipath effect, and the penetration loss simultaneously.

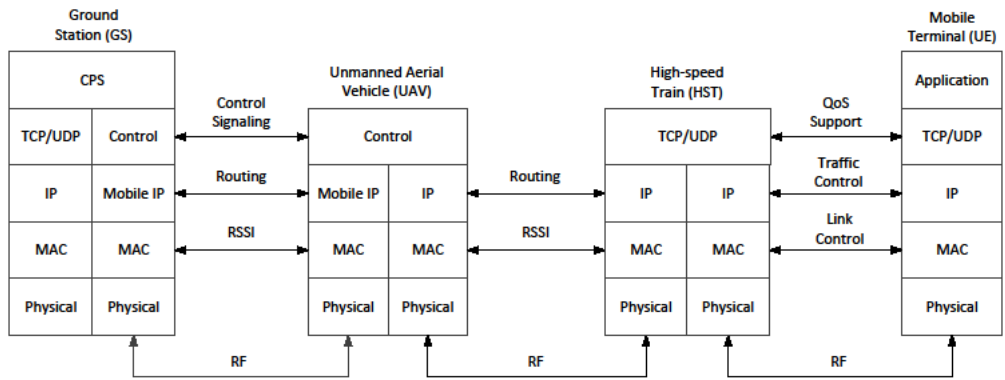

**Fig. 3**　The three-hop layer model

## 3.3 Communication types

As is illustrated in Fig. 2, based on the elements involved in the UAS-R communication system, there are three types of direct communications, i.e. communications between GS and UAV, communications between UAVs, and communications between HST and UAV. These communications are necessary for interactions between the UAS-R system elements to form a network that can sense and control the physical world. In the heterogeneous UAS-R system, satellites can assist communications in a long range. Therefore, communications between satellite and GS or UAV may also be possible and necessary, and it is widely discussed in satellite communications. In this section, the former three types of communications will be analyzed in detail.

### 3.2.1 Air-to-Ground communications

Naturally, the cellular infrastructure is not designed for the Air-to-Ground (A2G) communications. Standard cellular network channel models focus on the terrestrial coverage and do not provide a sufficiently accurate vertical channel characterization. Fast fading effects and Doppler shift caused by mobility have less impact on UAV communications compared with cellular communications [7].

The A2G links are established between the UAVs and the GSs, and are used for bidirectional transmission of control data, measurement data, and user data. In this scenario, the GS is no longer a conventional BS. The GS covers a much larger area than a femto-cell which is typically used in the LTE (Long Term Evolution) system. Actually, the GS does not represent a large cell. It applies smart antennas and power control to target the UAVs via beamforming. At the same time, the UAV uses beamforming for source GS separation. Therefore, there is no cell concept for the GS. However, a narrow fan sector is created in the air. The UAVs are generally located in these sectors.

The UAV provides a solution to the problems encountered in high-speed railway scenarios. Because the altitudes of UAVs allows them to jump out of the terrestrial shadow and to obtain the line-of-sight (LoS) or near LoS (NLoS) communication channels over a large area, i.e. this approach eliminates most of the multipath and shadow effects. The foremost characteristic of A2G channels is that the LoS is often dominant. Thus, they have a tendency to encounter much less severe fading and shadowing, less path loss and shadowing variation compared with terrestrial channels. Even the height of a UAV can introduce more path loss, but the gain of a smart directional antenna can compensate this loss. In [8], the results indicate that a significant multipath effect was observed at low angles (e.g. 7.5°) while high angle (e.g. 30°) channels have negligible multipath response. In the UAS system, UAVs are usually at angles range from 10° to 45°. Hence, the multipath response is negligible.

In free space wireless communication scenarios, interferences are everywhere. The interferences not only cause high packet error rates (PER), low throughputs, and low signal-to-noise-plus-interference ratio (SINR), but also increase the round trip times (RTT) and jitters of the system. In addition, to support high QoS and high safety assurance, interruption-free connectivity is indispensable. To achieve an interruption-free and seamless connection, the UAVs will perform handovers between different GSs as needed during their flights. A connection will be interrupted when the signal from the source station is unable to meet the minimum communication requirements, e.g. receiver's sensitivity, minimum SINR, or minimum Received Signal Strength Indicator (RSSI). If the signal from the target/next GS is strong enough to establish a connection at the same time, then a handover is triggered. For example, the UAV continuously measures the RSSI values in an averaged update interval of 250 ms. Each UAV moves along a certain linear path. When a certain RSSI-threshold is reached, it indicates that the maximum transmission range is almost reached. Simply using the RSSI as a metric, channel dynamic effects, such as noise, interference and high outage probability can be mitigated. However, schemes and algorithms are desired to solve and optimize the interference and handover problems.

### 3.2.2  Air-to-Train communications

In the Air-to-Train (A2T) communications, the UAVs act as airborne relays between the HSTs and the GSs for the purpose of providing high-speed data link and seamless connectivity for train users, i.e. train passengers and train operators.

The A2T and A2G channels share many characteristics, such as less shadowing effect, less multi-path effect, and LoS or NLoS dominates the channel [9]. The major difference is that the A2T channel can almost eliminate the Doppler Effect owing to no relative moving between the UAV and the HST, under the assumption that the UAV has the exact same speed with the HST and flies at the same height. Even if this assumption is not always valid, the variation is trivial through intelligent operation of the UAV. Therefore, the A2T channel is approximated as a stationary point-to-point model. Considering that the distance between the HST and the UAV is nearly the height of the UAV (e.g. at the order of 1 km), the millimeter wave technology can be applied to provide broadband communications. Millimeter wave (30-300GHz) has a strong directivity, so that the mutual signal interference is very small. In addition, millimeter wave antennas are more easily to be miniaturized and integrated in an onboard system.

In the terrestrial cellular network designs, Code Division Multiple Access (CDMA) or Space Division Multiple Access (SDMA) [10] is an approach to increase the capacity of the network where Time Division Multiple Access (TDMA) or Frequency Division Multiple Access (FDMA) is already applied. Unlike terrestrial cellular network designs, the CDMA or SDMA approach is considered to solve the separation problem for those HSTs that have been allocated to same time or frequency resources. However, different orthogonal resources can be allocated to different HSTs through CDMA or SDMA method.

For instance, since the HSTs in the same track are and shall be long distance separated, the bandwidth is divided into two equal-sized portions. One portion of the bandwidth is assigned to the HSTs moving in the same direction (there are usually two directions). Same frequency band is used to send data to one HST in an Aerial Moving Cell provided by associated UAV. The AMC always moves along with the HST and envelops this HST for most of the time. In case of hard handovers or flight changes, the connections may be dropped. Flight changes mean that a UAV is reaching its flight range or endurance limit, so that an alternative UAV is on the way to proceed with the former's work. Usually, flight changes happen when the HST is at a train station. There exist some wireless networks, such as 3G, LTE, or Wi-Fi, near the train station. Thus, the HST can perform vertical handovers to any available network seamlessly, because its speed is reduced in order to stop at the station. From an overall system point of view, only one frequency band is used for one HST. Each HST can achieve broadband communications and seamless handovers. The utilization of UAVs not only enhances the system capacity, but also reduces the outage probability. Therefore, this kind of approach will improve the safe operation of the HSTs and communication QoS of the passengers.

### 3.2.3 Air-to-Air communications

The Air-to-Air (A2A) channels are viewed as free space channels based on the assumption that a dominant LOS path is available [11]. When two HSTs are running in opposite directions on adjacent tracks respectively or are running closely in the same track, two UAVs may fly near each other. In the former scenario, the UAVs can gain many benefits, e.g. collision avoidance, from the A2A communications. They exchange information about position, altitude, recently identified ground obstacles, and flight path from each other. In the latter scenario, the A2A communications help to control the HST operation to avoid possible accidents like rear-end collision without the help of the GS or the train control center. The idea of this assistance is that a source UAV (associated with source HST) uses another UAV (associated with target HST) within its communication range as a relay to transmit signals to the target HST on the same track. Additional A2A links are used for relaying those UAVs that are during an outage or a connection failure to GSs within the communication range. Besides, the A2A links also allow both control data and user data to get routed through the A2A distributed network.

## 4 System feasibility analysis

The difference between the UAS-R system and other train control systems is the utilization of UAVs as air relay nodes which have the sense and control functions. Since the ground system is well developed, the system feasibility analysis mainly rests on the communication capabilities of the UAV.

### 4.1 UAV communication requirements

The UAVs require significantly more communications than conventional manned aircrafts. Roughly speaking, these communication requirements can be classified into link connection, remote control and monitoring, and route plan. Link connection mainly supports three kinds of channels: control signaling channel, data collecting and sending channel, and data forwarding channel. The control signaling channel is related to receiving operation commands from GSs and transferring operation commands to HSTs. The data collecting and sending channel is used to collect ground information and transfer the ground information to the GSs. The data forwarding channel is particularly used to forward user data between the HST and the GS. Remote control and monitoring is related to remote train operation control and detecting or remote sensing emergences. This requires onboard digital visual equipment and cooperative information sharing between the UAVs. Route Plan [12] has equivalent abilities of collision avoidance, flight optimization, and HST-tracking.

The average parameters of military and civil UAV and the estimated UAV capabilities in the UAS-R system are shown in Table 1. These parameters provide a rough impression about the three different applications of specific systems. In terms of delay, the one-way propagation delay is less than 0.2 ms, the UAV data processing time is less than 20 ms, the HST data processing time is about 40 ms (which is an over-estimation considering the HST will integrate and extract data stream), and then the overall delay from the GS to end user is less than 61 ms.

**Table 1** UAV parameters

| Parameters | Military | Civil | UAS-R |
| --- | --- | --- | --- |
| Endurance (hour) | > 20 | < 5 | 1-5 |
| Range (km) | 300 | < 10 | 50-100 |
| Altitude (m) | 600-5000 | 25-200 | 100-500 |
| Payload (kg) | 200 | 5 | 20 |
| Speed (km/h) | 30-440 | < 25 | 250-350 |
| Data Rate (Mbps) | 10-274 | 10 | > 1000 |
| Band | L/C/X/Ku/Ka | UHF/X | UHF/X/Ka/U |

**4.2 UAV functionalities**

In wireless networks, airborne nodes serve as mobile broadband storage points, or serve as relay nodes between ground mobile devices and ground stations or satellites, to support seamless communications. The railway status and environment information is the foundation of train operation and management. Current methods of collecting the information are stationary, because the sensors are installed in fixed locations. In addition, stationary sensors are insufficiently deployed due to the high costs of purchase, installation, and maintenance. Therefore, the railway status and environment information is almost unattainable. Compared with conventional relay devices and stationary sensors, the UAVs have numerous advantages such as high mobility, high flexibility, low cost, and extensive and profound visions.

As is shown in Fig. 4, through a high-speed data bus, all onboard embedded subsystems are connected. Typically, higher computational capabilities of the central processing unit (CPU) and other collaborating controllers are required in the UAV's embedded subsystem. The control manager, as one of the onboard subsystems, is responsible for initialization and maintenance of the communication links with all potential possibilities, making decisions about when and how to route or forward data, and providing QoS for different applications or missions. All the communication devices in the UAS-R system are connected to the UAV's embedded subsystem via wireless interfaces. For the A2G and A2T links, the LTE technology can be utilized. In terms of telematics, all the above communication devices are synchronized to exchange control and system status data. It is necessary to introduce gateway functions which also allow information exchange among diverse communication entities (e.g. eNB, satellite, data bus, etc.) and protocols. Mobile broadband storage point is ensured by a large memory or cache array. The storage management is capable to read and write large amount of data needed in transmission and cache ongoing data during handovers or link failures. Software Defined Radio (SDR) subsystem is associated with smart antenna control subsystem. Both of these subsystems are recognized as key technologies of the UAVs.

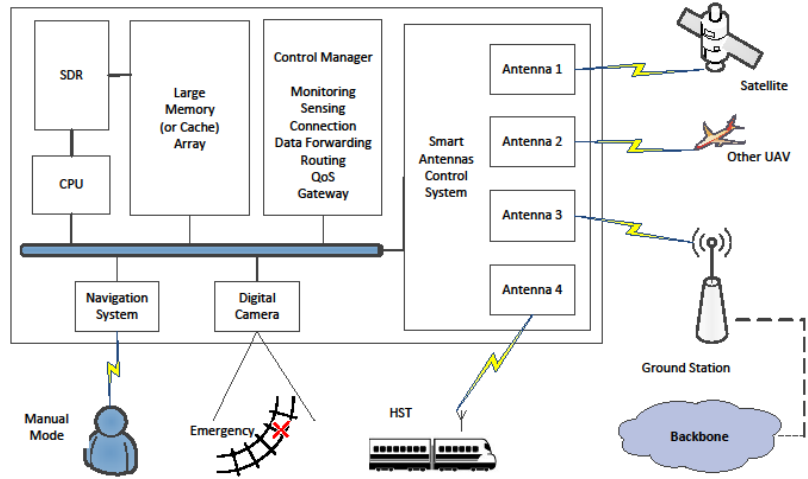

**Fig. 4**  UAV function descriptions

In addition, the UAV equips with various customized sensors, such as optical, thermal, infrared, and distance sensors, to capture or measure various data interested, no matter it is day or night, rainy or foggy which may be hazardous to the HST safety. For example, each image or video captured by a high-resolution digital camera is sent to the GS in real time. The GSs or UAVs utilize the data to infer the real environments of the physical world based on coordinates and time stamps obtained from the onboard navigation system (e.g. GPS). In the near future, the image progressing technique will be upgraded to autonomous generate real-time 3D virtual realities of the environment. Much other assistant information such as train velocity and weather condition can be obtained from special sensors respectively.

**4.3  Trade-offs**

The construction of long range and high-speed transmission communication system often accompanies with risks and difficulties. For example, on the one hand, when the distance is doubled, the transmission power or antenna area must be almost tripled in order to maintain the equivalent transmission rate. On the other hand, when the transmission rate is doubled, the transmission power or antenna area must be roughly doubled as well. Special attentions should be paid to the payload, power supply, and heat-sinking capability, size, and cost of the UAVs. For the high-speed railway communication requirements, the UAS-R has greater communication needs, such as throughput, outage probability, and latency or delay, than the current manned aircrafts and the cellular systems. In addition, current approaches that apply powerful long-range communications or satellite communications are too expensive and large-sized for small aircrafts. While smaller radios fundamentally limit the UAV performances in terms of range, altitude, and payload. Besides, the UAV communication quality suffers higher losses in adverse meteorological conditions. When the train is located in a tunnel, the connections must transfer to substitute communication facilities.

**4.4 Key technologies**

Advancements have been made on many fronts, including mechanical engineering, dynamics, energy storage, and autonomous control. However, great efforts still should be made to fulfill the stringent UAS-R communication requirements. Some physical (PHY) layer and media access control (MAC) layer challenges are enumerated as follows:

1) Airborne dynamics induce signal variations;

2) Several ground level multipath environments;

3) Partial or full blockage of the LOS path;

4) Interference from other terrestrial RF sources;

5) Multiple access and resource allocation of UAVs and HSTs.

In addition, some key technologies which will strongly affect the system performance and could iron out the above challenges are introduced in this subsection.

**4.4.1 Antenna technology**

The development trend of antenna technology is intelligent, functional diversification, and massive-scale. Smart Antennas [13] is one kind of the intelligent antenna technologies. The smart antennas arrays applied with smart signal processing algorithms can identify spatial signals such as the direction of arrival (DOA), perform beamforming, and track and locate the mobile devices. The GS employs smart antennas to track the UAVs. Similarly, the UAV employs multi-band smart antennas to track the HSTs, GSs, satellites, and other UAVs. Through the SA technology, the system can track and communicate high-speed mobile objects. Therefore, the control and monitor of the HST and UAV are feasible. The functional diversification of antennas can be realized by multi-band and multi-mode antennas. The massive-scale effect of the antennas is achieved via deploying multiple antennas, i.e. Multiple Input Multiple Output (MIMO) model [14]. MIMO system exploits the multipath propagation phenomena rather than eliminates the multipath effects, to increase system throughput (i.e. multiplex gain) and to reduce the bit error rate (i.e. diversity gain). Given these MIMO system benefits, the UAS-R system also considers the multi-band and multi-mode communication requirements required by different communication scenarios. At the GS-side, UAV-side, and the HST-side, multiple transmitters and receivers are available. Therefore, the UAS-R system can set up A2G, A2A, and A2T communication links adaptively and achieve diversity and multiplex gain. Efficient communications can be achieved. However, MIMO transceivers may be more complex, heavier, and larger compared with single transceivers. And the requirements of fabricating antennas components for the UAV are stringent and difficult.

### 4.4.2  Software defined radios

Generally, SDR [15] systems consist of some pre-implemented hardware, such as processor, storage, analog-to-digital signal converter, RF front end, etc., and several dedicated software. This design enables a device to efficiently transmit and receive various standards wireless signals based on the software that is currently used. Therefore, the SDR system has significant utility for services that involve a wide variety of real-time changing wireless protocols. Through the onboard programmable imbedded subsystems, the UAVs can switch operation mode between different kinds of services or missions smoothly. However, the hardware computational capabilities and the effectiveness of software algorithms are technical bottlenecks for the

UAVs. More efficient and reliable micro devices and algorithms to support the UAS-R communications are still under research or test.

### 4.4.3    Joint modulation and coding

Many modulation techniques, e.g. continuous phase modulation (CPM), Gaussian minimum-shift keying (GMSK) modulation, and Orthogonal Frequency Division Multiplexing (OFDM) modulation, can be applied to the UAS-R system. Particularly, multi-carrier modulation schemes can provide the system required robustness and throughput performances. Different applications utilize different modulation schemes. Therefore, to combine various modulation schemes through the automatic modulation classification (AMC) [16] is a way to joint modulations. The AMC technology can switch between signal detection and demodulation intermediately, and adapt to different wireless channel conditions to support communications.

The Flexible Low Density Parity Check (F-LPDC) [17] coding with compressed sensing (CS) [18] is very suitable for the proposed UAS-R system. The F-LPDC code is a type of linear block codes approaching the Shannon limit. It can be constructed with sparse parity-check matrixes and decoded with highly parallelizable decoder architectures. Compressed sensing takes advantage of the sparseness and compressibility of the F-LPDC coding signals on the condition that the sampling rate is much less than the Nyquist sampling rate. Through signal random sampling and solving an underdetermined linear problem, the entire signal can be reconstructed from relatively fewer measurements. Since CS is effective in compression of the samplings, the information content is concentrated and the transmission time is shortened obviously. F-LPDC coding with CS increases the efficiency of data transmissions, thereby ideally in accordance with the broadband and real time communication requirements of the UAS-R system.

### 4.4.4    Multiple access

Multiple access techniques are widely applied in terrestrial cellular systems as MAC layer protocols.

However, spatial multiple access is still feasible and may be beneficial to the proposed UAS-R system. The key issue of access techniques is the orthogonal resource allocation to avoid interferences and collisions.

In the SDMA scheme, the UAV utilizes beamforming to separate HSTs and to mitigate interferences, achieving spatial diversity gain. The UAV tracks the HST via position predictions and targets its beam at the HST to maximize the SINR at each time slot. It is assumed that at time step $n$-1, all the HSTs feedback their current speed and position information (available in train control center) to the UAV. Then the information is utilized to predict the HST positions at time step $n$. Since there is no interference from other users under a TDMA scheme, this beamforming is equivalent to the maximum ratio combine method. The CDMA scheme is generally known as its efficient utilization of fixed frequency spectrum, flexible resource allocation, and spread-spectrum characteristics to avoid interference. For this scheme, each UAV or HST is assigned an identification code for data transmission, and the system uses predefined Walsh code sequences to distinguish and identify different HSTs and UAVs. In this case, channels are shared in order to avoid strong multi-access interferences, the GSs can track the operation of each UAV or HST, and the UAVs can also recognize each other and monitor the HSTs.

## 5   Summary

This article proposes a wireless communication system solution which provides broadband Internet access to high-speed train users in the future Cyber-physical systems. The solution is based on airborne relay to reduce the costs of many base stations and antenna units near the trackside. Meanwhile, an aerial moving cell concept is further proposed in order to support an always online connection for the high-speed train operators and passengers. In addition, required UAV functions are discussed in detail. The system architecture prototype including A2G, A2T, and A2A communications is proposed and analyzed. The feasibility analysis shows that using UAVs to support high-speed train operations and communications is possible and promising. More sufficient simulations and experiments will be undertaken in the future. However, issues such as wireless channel modeling, system protocol design, system performance analysis, efficient information extraction and fusion, and UAV flight scheduling and experiments, etc. still remain open and need further explorations. To realize the broadband communications of high-speed mobile users and zero accident of transportations, the proposed system solution is very promising in the future CPS systems.

**References**

1.  Sha, L., Gopalakrishnan, S., Liu, X., & Wang, Q. Cyber-Physical Systems: A New Frontier. In *2008 IEEE International Conference on Sensor Networks, Ubiquitous, and Trustworthy Computing, 2008* (pp. 1-9). doi:10.1109/sutc.2008.85.

2.  Frew, E. W., & Brown, T. X. (2008). Airborne communication networks for small unmanned aircraft systems. *Proceedings of the IEEE, 96*(12), doi:10.1109/JPROC.2008.2006127.

3.  D., C., Gavrilovich, J., Ware, G. C., & LLP, F. (2001). Broadband Communication on the Highways of Tomorrow. *IEEE Communications Magazine*(4), 146-154, doi:10.1109/35.917517.

4.  Lannoo, B., Colle, D., Pickavet, M., & Demeester, P. (2007). Radio-over-Fiber-Based Solution to Provide Broadband Internet Access to Train. *IEEE Communications Magazine*(February), 56-62, doi:10.1109/MCOM.2007.313395.

5.  White, G. P., & Zakharov, Y. V. (2007). Data Communications to Trains From High-Altitude Platforms. *IEEE TRANSACTIONS ON VEHICULAR TECHNOLOGY, 56*(4), 2253-2266, doi:10.1109/TVT.2007.897185.

6.  Yiqing, Z., Zhengang, P., Jinlong, H., Jinglin, S., & Xinwei, M. Broadband wireless communications on high speed trains. In *Wireless and Optical Communications Conference (WOCC), 2011 20th Annual, 15-16 April 2011 2011* (pp. 1-6). doi:10.1109/wocc.2011.5872303.

7.  Feng, Q., McGeehan, J., Tameh, E. K., & Nix, A. R. Path loss models for air-to-ground radio channels in urban environments. In *Vehicular Technology Conference, 2006. VTC 2006-Spring. IEEE 63rd,*

*2006* (Vol. 6, pp. 2901-2905): IEEE. doi:10.1109/VETECS.2006.1683399.

8. Newhall, W. G., Mostafa, R., Dietrich, C., Anderson, C. R., Dietze, K., Joshi, G., et al. Wideband air-to-ground radio channel measurements using an antenna array at 2 GHz for low-altitude operations. In *Military Communications Conference, 2003. MILCOM'03. 2003 IEEE, 2003* (Vol. 2, pp. 1422-1427): IEEE. doi:10.1109/MILCOM.2003.1290436.

9. Goddemeier, N., Daniel, K., & Wietfeld, C. (2012). Role-Based Connectivity Management with Realistic Air-to-Ground Channels for Cooperative UAVs. *Selected Areas in Communications, IEEE Journal on, 30*(5), 951-963, doi:10.1109/JSAC.2012.120610.

10. Bagadi, K. P., & Das, S. (2013). Efficient complex radial basis function model for multiuser detection in a space division multiple access/multiple-input multiple-output-orthogonal frequency division multiplexing system. *IET Communications, 7*(13), 1394-1404, doi:10.1049/iet-com.2012.0688.

11. Leitgeb, E., Zettl, K., Muhammad, S., Schmitt, N., & Rehm, W. Investigation in free space optical communication links between unmanned aerial vehicles (UAVs). In *Transparent Optical Networks, 2007. ICTON'07. 9th International Conference on, 2007* (Vol. 3, pp. 152-155): IEEE. doi:10.1109/ICTON.2007.4296268.

12. Jiang, F., & Swindlehurst, A. (2012). Optimization of uav heading for the ground-to-air uplink. *Selected Areas in Communications, IEEE Journal on, 30*(5), 993-1005, doi:10.1109/JSAC.2012.120614.

13. Viani, F., Lizzi, L., Donelli, M., Pregnolato, D., Oliveri, G., & Massa, A. (2010). Exploitation of parasitic smart antennas in wireless sensor networks. *Journal of Electromagnetic Waves and Applications, 24*(7), 993-1003, doi:10.1163/156939310791285227.

14. Rusek, F., Persson, D., Lau, B. K., Larsson, E. G., Marzetta, T. L., Edfors, O., et al. (2013). Scaling up MIMO: Opportunities and challenges with very large arrays. *Signal Processing Magazine, IEEE, 30*(1), 40-60, doi:10.1109/MSP.2011.2178495.

15. Mueck, M., Piipponen, A., Kalliojarvi, K., Dimitrakopoulos, G., Tsagkaris, K., Demestichas, P., et al. (2010). ETSI reconfigurable radio systems: status and future directions on software defined radio and cognitive radio standards. *Communications Magazine, IEEE, 48*(9), 78-86, doi:10.1109/MCOM.2010.5560591.

16. Dobre, O. A., Abdi, A., Bar-Ness, Y., & Su, W. (2007). Survey of automatic modulation classification techniques: classical approaches and new trends. *Communications, IET, 1*(2), 137-156, doi:10.1049/iet-com:20050176.

17. Zhang, C., Wang, Z., Sha, J., Li, L., & Lin, J. (2010). Flexible LDPC Decoder Design for Multigigabit-per-Second Applications. *IEEE Transactions on Circuits and Systems I-Regular Papers, 57*(1), 116-124, doi:10.1109/tcsi.2009.2018915.

18. Xiang, L., Luo, J., & Rosenberg, C. (2013). Compressed Data Aggregation: Energy-Efficient and High-Fidelity Data Collection. *IEEE-ACM Transactions on Networking, 21*(6), 1722-1735, doi:10.1109/tnet.2012.2229716.